\documentclass[grl]{agu2001}
\usepackage[dvips]{graphicx}
\lefthead{HELMSTETTER and SORNETTE}
\righthead{Importance of triggered seismicity}	
\authoraddr{Agn\`es Helmstetter,  Institute of
Geophysics and Planetary Physics, University of California, Los Angeles,
California.  (e-mail: helmstet@moho.ess.ucla.edu)}
\authoraddr{Didier Sornette,
Department of Earth and Space Sciences and Institute of
Geophysics and Planetary Physics, University of California, Los Angeles,
California and Laboratoire de Physique de la Mati\`{e}re Condens\'{e}e,
CNRS UMR 6622 and Universit\'{e} de Nice-Sophia Antipolis, Parc Valrose,
06108 Nice, France (e-mail: sornette@moho.ess.ucla.edu)}

\input{update.tex}
\begin{document}
\title{Importance of direct and indirect triggered seismicity
in the ETAS model of seismicity}
\author{Agn\`es Helmstetter$^1$ and Didier Sornette$^{1,2,3}$}
\affil{$^1$ Institute of
Geophysics and Planetary Physics, University of California, Los Angeles,
       California 90095-1567\\
$^2$ Department of Earth and Space Sciences, University of California, 
Los Angeles,       California 90095-1567 \\
$^3$    Laboratoire de Physique de la Mati\`{e}re Condens\'{e}e, CNRS UMR 6622
Universit\'{e} de Nice-Sophia Antipolis, Parc Valrose, 06108 Nice, France}

\newcommand{\be}{\begin{equation}}
\newcommand{\ee}{\end{equation}}
\newcommand{\ba}{\begin{eqnarray}}
\newcommand{\ea}{\end{eqnarray}}
\newenvironment{technical}{\begin{quotation}\small}{\   end{quotation}}


\begin{abstract}

Using the simple ETAS branching model of seismicity, which assumes 
that each earthquake can trigger other earthquakes, we quantify
the role played by the cascade of triggered seismicity in
controlling the rate of aftershock decay as well as the overall level
of seismicity in the presence of a constant external seismicity source.
We show that, in this model, the fraction of earthquakes in the 
population that are aftershocks is equal to the fraction of aftershocks 
that are indirectly triggered and is given by the average number of 
triggered events per earthquake. Previous observations that a significant 
fraction of earthquakes are triggered earthquakes therefore imply that 
most aftershocks are indirectly triggered by the mainshock.
\end{abstract}

\begin{article}

\section{Introduction}
There is a growing awareness and an intense research activity based on
the fact that a significant fraction of earthquakes are
events triggered (in part) by preceding events. In addition, a significant
part of triggered events may be indirectly triggered by a previous event
through a cascade process. What is then the relative
role of earthquake interactions and triggering compared with the
underlying tectonic driving forces? Is there a way
to distinguish triggered earthquakes from untriggered ones
or to estimate the proportion of directly or indirectly triggered 
earthquakes? Here, we use the Epidemic-Type Aftershock Sequence
(ETAS) model to offer a quantification of earthquake interactions.
This model is based on the two best established empirical laws of 
seismicity, the Gutenberg-Richter and the Omori law.
The ETAS model has been used in many studies to describe or predict
the spatio-temporal distribution of seismicity and reproduces
many properties of real seismicity (see [{\it Ogata}, 1999] 
and [{\it Helmstetter and Sornette}, 2002] for reviews). 
The ETAS model assumes that the seismicity results from the sum of 
an external constant loading  and from  earthquakes triggered 
by these sources in direct lineage or through a cascade of generations.
From this definition (see below), it is clear that the ETAS model
is not only a model of aftershock sequences, as the acronym ETAS
would make one to believe, but describes the global seismicity
including background and interacting triggered seismicity.
We use this model to quantify (a) the fraction of triggered events 
relative to the sources and (b) the fraction of indirectly triggered 
events with respect to the total triggered seismicity.

Question (a) has been previously visited in order to provide
unambiguous definitions of aftershocks and to decluster seismic 
catalogs. Several alternative algorithms
for the definition of aftershocks have been proposed
[see {\it Molchan and Dmitrieva}, 1992 for a review].
{\it Gardner and Knopoff} [1974]  and  {\it Knopoff} [2000] 
used a  windowing method and
found that 2/3 of the events in the catalog of Southern California
 are aftershocks. {\it Reasenberg} [1985] analyzed the
central California catalog and  found that
48\% of the events belong to a seismic cluster.
{\it Davis and Frohlich} [1991] used the ISC catalog and found that 
30\% of earthquakes belong to a cluster, of which 76\% are aftershocks 
and 24\% are foreshocks. {\it Kagan} [1991] estimated the ratio of 
dependent events in various catalogs (California and worldwide) using 
an inversion by the maximum likelihood method of the ETAS model.
The proportion of dependent earthquakes of the first generation that he
estimated displays huge fluctuations from 0.1\%  for deep events to
90\%, but is often close to 20\%.

With respect to question (b), it has long been suggested that 
aftershocks may produce their own aftershocks, commonly known
as secondary or indirect aftershocks.
The observation of large and sudden changes of the seismicity rate 
after a mainshock  [e.g. {\it Correig et al.}, 1997] and the 
existence of strong spatio-temporal clustering of aftershocks 
shows that a significant proportion of aftershocks may be
triggered indirectly by the mainshock, that is, they may be aftershocks
of aftershocks triggered by the mainshock [{\it Felzer et al.}, 2003].
For instance in Southern California,
the $M=6.5$ Big-Bear earthquake occurred a few hours following
the Landers $M=7.3$ event and has clearly triggered its own
aftershock sequence. While each aftershock induces a negligible
stress change by comparison to the mainshock, all aftershocks when 
taken together can significantly alter the stress field induced by 
the mainshock, so that most aftershocks at large times after the 
mainshock are triggered by previous aftershocks of the mainshock.
{\it Felzer et al.} [2002] estimated the rate of indirect aftershocks,
from a comparison of the Landers aftershock sequence with numerical
simulations of the ETAS model. They found that
about 85\% of the aftershocks of the Landers event were indirect
aftershocks. This implies that the 1999 $M_W=7.1$ Hector Mine
earthquake was triggered, not by the  1992 $M_W=7.3$ Landers earthquake 
itself [{\it Felzer et al.}, 2002], but more likely by some of its direct
and indirect aftershocks. {\it Felzer et al.} [2003] further analyzed the 
temporal evolution of the proportion of secondary aftershocks.
They found that, after a few days or weeks following
a mainshock depending on mainshock magnitude, most aftershocks are 
secondary aftershocks.
We now recall the formulation of the ETAS model and its main results
on the importance of triggered seismicity.

\section{The ETAS model of triggered seismicity}

The present parametric form of the ETAS model used in this paper
was formulated by {\it Ogata}  [1988]. We refer to  [{\it Ogata}, 1999; 
{\it Helmstetter and Sornette},  2002] for reviews on the ETAS model 
and for a discussion of the model parameters.
The ETAS model assumes that a given event of magnitude $m_i \geq m_0$ 
occurring at time $t_i$ triggers other events in the time interval
between $t$ and $t+dt$ at the rate
\be
\phi_{m_i}(t-t_i) = \rho(m_i) ~\Phi(t-t_i)~.
\label{eq1}
\ee
$\Phi(t)$ is the direct Omori law normalized to $1$
\be
\Phi(t) = {\theta~c^{\theta} \over (t+c)^{1+\theta}},
\label{Phi}
\ee
where $c$ is a regularizing time scale that ensures that the seismicity 
rate remains finite close to the mainshock.
The average  number of aftershocks triggered directly
by an event of magnitude $m$ is 
\be
\rho(m)=k~10^{\alpha (m-m_0)}~,
\label{rho}
\ee
where $m_0$ is a lower bound magnitude below which no daughter is triggered.
The model is complemented by assuming that each earthquake has a magnitude
 independently chosen according to the density distribution $P(m)$.
The magnitude distribution is usually taken equal to the Gutenberg-Richter 
law $P(m) \sim 10^{-b(m-m_0)}$  with eventually a cut-off 
for large magnitudes. The model can also be extended to
include the spatial distribution of seismicity [{\it Ogata} 1999].
The key parameter of the ETAS model (\ref{eq1}) is the average number
(or ``branching ratio'') $n$ of directly triggered earthquakes
per mother-event. This average is performed over time and over
all possible mother magnitudes. The branching ratio has a finite value 
for $\theta>0$ equal to
\be
n \equiv \int \limits_0^{\infty} dt
\int \limits_{m_0}^{\infty}~P(m)~\rho(m)~\Phi(t)~dm~.
\label{nvaqlue}
\ee
The normal regime corresponds to the subcritical case $n<1$ for which
the seismicity rate decays after a mainshock to a constant level (in
the case of a steady-state source). Note that the realized number 
of aftershocks for a given earthquake is not $n$ but depends on its magnitude,
 according to the function $\rho(m)$ given by (\ref{rho}).

The total seismicity rate (or intensity) $\lambda(t)$ at time $t$
is given by the sum of the ``external'' source $s(t)$
and of the aftershocks triggered by all previous events 
\be
\lambda(t)=s(t)+\sum_{i|t_i \leq t} \phi_{m_i}(t-t_i)~.
\label{lambda}
\ee
This external source $s(t)$ acts as an external driving force 
ensuring that the seismicity does not vanish.

Taking the ensemble average of (\ref{lambda}) over many possible
realizations of the seismicity, we obtain the following equation
for the first moment or statistical average $N(t)$ of $\lambda(t)$
[{\it Sornette and Sornette}, 1999; {\it Helmstetter and Sornette}, 2002]
\be
N(t)=s(t)+ n \int \limits_{-\infty}^{t} ~\Phi(t-\tau)~N(\tau)~ d\tau.
\label{N1}
\ee
The average seismicity rate is the solution of this self-consistent
integral equation, which embodies the fact that each event may start a
sequence of events, which can themselves trigger secondary events, and so on.

The global rate of aftershocks including indirect aftershocks
triggered by a mainshock of magnitude $M$ occurring at $t=0$ is given by
$\rho(M) K(t)/n$, where the renormalized Omori law $K(t)$ is obtained as a
solution of (\ref{N1}) with the general source term $s(t)$ replaced by
the Dirac function $\delta(t)$. The solution for $K(t)$ is given in 
[{\it Helmstetter and Sornette}, 2002] and is illustrated in Figure 
\ref{figKphi}. The effect of the cascade of direct, secondary, and
later-generation aftershocks is to renormalize the bare Omori law 
$\Phi(t) \sim 1/t^{1+\theta}$ into $K(t) \sim  1/t^{1-\theta}$ at 
early times $t \ll t^*$ where $t^* \approx c |1-n|^{-1/\theta}~$. 
The characteristic time $t^*$ 
is infinite for $n=1$ and becomes very small for $n \ll 1$.
Figure \ref{figKphi} also shows the rates $N_i(t)$ of
aftershocks of generation $i$, for $i=1$ to $20$. Taking an ensemble 
average, we predict $N_1(t) = \rho(M) \Phi(t)$,
$N_2(t) = \int_0^t n \Phi(t-\tau)~ \rho(M)~\Phi(\tau)  d\tau$, and
more generally 
\be 
N_i(t) = n \int_0^t \Phi(t-\tau)~ N_{i-1}(\tau)~ d\tau~,
\label{Ni}
\ee
such that the total seismicity rate is reconstructed as the sum
$N(t) =\sum_{i=1}^\infty N_i(t)$. Figure \ref{figKphi} illustrates
clearly the role and importance of the successive generation of indirect
aftershocks in the construction of the global observable seismicity.

In real data, it is impossible to distinguish unambiguously
aftershocks from background seismicity, or direct aftershocks from
indirect aftershocks. The distinction is only probabilistic. 
Each event results in part from the external loading and in part
from the effect of all previous earthquakes. Knowing the parameters
of the model, we can however estimate the probability that each event
results from the external source or is an aftershock of a previous
earthquake [{\it Kagan}, 1991].
In the sequel, we estimate the ratio of triggered seismicity  over total
seismicity in section \ref{sn1} and the proportion of secondary aftershocks  
over total aftershocks in section \ref{sn2}, and we show that these two 
quantities are equal to the branching ratio $n$.

\section{Proportion of aftershocks}
\label{sn1}
Let us consider the situation in which $s(t)$ corresponds to a constant
Poisson source process with intensity $\mu$, representing the effect of
the external loading. Then, the observed seismicity results both from this
constant source rate and from the direct and indirect
aftershocks triggered by this constant external loading.
In the regime $n<1$, the global seismicity is stationary, with large
fluctuations following large earthquakes due to the triggered
aftershock sequences. The rate of aftershocks $r_0$ triggered directly
by the tectonic source $\mu$ is on
average $r_1=\mu n$ because each single  event triggers on average
$n$ events, when averaging over all magnitudes.
The rate of second generation aftershocks, triggered by aftershocks of the
tectonic source, is $r_2=n r_1=\mu n^2$.
At the $i^{th}$ generation, the rate of aftershocks triggered indirectly by
the tectonic source $\mu$ is given by $r_i=\mu n^i$.
Summing over all generations, the global rate $R_{aft}$ of direct and
indirect aftershocks of the constant external source in the sub-critical
regime $n<1$ is given by
\be
R_{aft.}=\sum_{i=1}^{i=\infty} r_i= \mu \sum_{i=1}^{i=\infty}
n^i={\mu n \over 1-n}~.
\label{S1}
\ee
The global seismicity rate $R$ is given by the sum of the
external loading $\mu$ and of the rate of aftershocks $R_{aft.}$:
\be
R=\mu+R_{aft.}= \mu+ {\mu n \over 1-n}={\mu \over 1-n}~.
\label{R1}
\ee
The result (\ref{R1}) shows that the effect of the cascade of aftershocks
of aftershocks and so on is to renormalize the external constant source
$\mu$ to a higher level $R$ that  increases as $n$ is close to the 
critical value 1, as illustrated in Figure (\ref{figmu}).
  This result is well-known in the branching process literature
[{\it Harris}, 1963] and has also been derived by {\it Kagan}  [1991]
for the slightly modified version of the ETAS model using $c=0$
and replacing it by an abrupt cut-off at early times.

The proportion of aftershocks (of any generation) is thus equal to
${R_{aft.} / R}=n$.
This expression shows that the average branching ratio $n$
can be directly observed from a suitable analysis of seismicity catalogs.
Indeed, clustering algorithms for detecting and
counting aftershocks provide a direct estimation and in general
a lower bound of $n$ because most triggered events cannot be distinguished
from the background seismicity. Note that the result (\ref{R1}) can
also be derived directly from the master equation (\ref{N1}) by inserting 
$s(t)=\mu$ in (\ref{N1}) and taking the expectation of $N(t)$.

\section{Proportion of indirect aftershocks}
\label{sn2}

There is another interpretation for $n$ as well as an additional
empirical tool to estimate it. We calculate the total number of
aftershocks $n_t$ triggered by a mainshock of magnitude $M$, including all
the generations of direct and indirect aftershocks, as follows.
The number of direct aftershocks is given by $n_1=\rho(M)$ using the
definition (\ref{eq1}).
The average number of second generation aftershocks $n_2$ is given by
the product of $n_1$ with the average number of aftershocks per earthquake
defined by $n$. Therefore $n_2=\rho(M)n$. The number of third
generation aftershocks of the mainshock is $n_3=\rho(M) n^2$.
The number of aftershocks for the $i^{th}$ generation is $n_i=\rho(M) n^{i-1}$.
The total number of aftershocks triggered by a mainshock of magnitude $M$
is thus given by
\be
S=\sum_{i=1}^{\infty} n_i=\rho(M)\sum_{i=0}^{\infty} n^i={\rho(M)\over 1-n}~.
\label{nt}
\ee
For $n \ll 1$, $S \approx \rho(M)$, i.e., most aftershocks are
directly triggered by the mainshock.
For $n \approx 1$, $S \gg \rho(M)$, i.e., most aftershocks are indirect
aftershocks of the mainshock.
The proportion of indirect aftershocks is given by
\be
{S-n_1 \over S}=
{{\rho(M) \over 1-n} -\rho(M) \over {\rho(M) \over 1-n}}=n~.
\label{r}
\ee
This result (\ref{r}) shows the fraction among all aftershocks of the
aftershocks triggered indirectly by the mainshock is given by the average 
branching ratio $n$, independently of the mainshock magnitude $M$.
We can also derive the result (\ref{r}) from the master equation
(\ref{N1}). Inserting $s(t)=\delta(t)\rho(M)$ in (\ref{N1}) and taking the
integral of (\ref{N1}) gives after some manipulation the global number 
of direct and indirect  aftershocks
$$
S = \int \limits_0^{\infty} N(t)dt
= \rho(M)+ n \int \limits_0^{\infty}~N(\tau)~d\tau
=\rho(M) +n ~S~,
\label{S2}
$$
which recovers expression (\ref{nt}) for $S$.

The branching ratio $n$ gives the proportion of indirect aftershocks
averaged over the whole aftershock sequence. It is different from
the instantaneous proportion of indirect aftershocks $\nu (t)$
that is defined by
\be
\nu(t) = {K(t) -\Phi(t) \over K(t)}  ~,
\label{nut}
\ee
which can be computed analytically using the expression of $K(t)$
given by {\it Helmstetter and Sornette} [2002].
The instantaneous proportion of indirect aftershocks increases from
0 for very small times $t \ll c$ (all aftershocks are triggered directly
by the mainshock) to a maximum value smaller than one at large times
$t \gg t^*$ given by
\be
\nu _{\infty} = \lim_{t \to \infty} \nu(t)
= 1 - (1-n)^2 ~{\theta~ \Gamma(\theta) \over  \Gamma(1-\theta)}~.
\label{nuinf}
\ee
The temporal evolution of $\nu(t)$ given by (\ref{nut}) is 
illustrated in the inset of Figure \ref{figKphi}. 

\section{Conclusion} 
We have shown that, in the ETAS model, the proportion of earthquakes 
that are triggered is equal to the proportion of aftershocks that 
are indirect, and is given by the branching ratio. 
 Previous observations that a significant 
fraction of earthquakes are triggered earthquakes therefore imply that 
most aftershocks are indirectly triggered by the mainshock.
 The importance of indirect aftershocks casts doubts 
on the relevance of prediction of aftershocks rate based on the
calculation of the Coulomb stress change induced by the mainshock
only, neglecting the stress changes induced by aftershocks  
[{\it Stein}, 1999]. It also opens the road for improved methods 
of seismicity forecasts [{\it Felzer et al.}, 2003].

\begin{acknowledgments}
We thank K. Felzer for useful discussions and for giving us
the code used to generate synthetic catalogs with the ETAS model.
We thank J.-R. Grasso and T. Gilbert for a careful reading
of the manuscript and useful discussions.
This work is partially supported by NSF-EAR02-30429, by
the Southern California Earthquake Center (SCEC) and by
the James S. Mc Donnell Foundation
21st century scientist award/studying complex system.
\end{acknowledgments}

{}

\end{article}

\begin{figure}
\includegraphics[width=8cm]{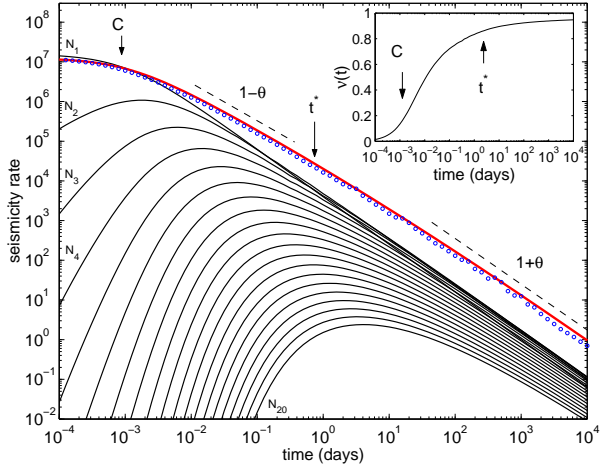}
\caption{\label{figKphi}
A realization of the ETAS model showing the realized seismicity rate 
$\lambda(t)$ (circles) following a $M=7$ mainshock
obtained by averaging over 1000 simulations
with  $n=0.8$, $\alpha=0.8$, $b=1$, $\theta=0.2$, $m_0=0$, $c=0.001$ day,
and the average renormalized propagator $K(t)$ (solid gray line).
The bell-shaped curves show the seismicity rates $N_i(t)$ of
aftershocks of generation $i$ estimated from equation (\ref{Ni}), 
for $i=1$ to $20$ from top to bottom.
The inset gives the proportion of indirect aftershocks $\nu(t)$
evaluated by (\ref{nut}). After 15 minutes, most aftershocks are triggered
indirectly by the mainshock.
At large times $t \gg t^*$, the proportion of indirect aftershocks
goes to an asymptotic value of 0.97.}
\end{figure}

\begin{figure}
\includegraphics[width=8cm]{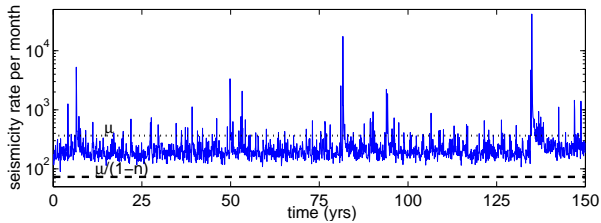}
\caption{\label{figmu} Rate of seismic activity for a synthetic catalog
generated using the ETAS model with parameters $\mu=0.1$ source events
per day, $c=0.001$ day, $n=0.8$, $\theta=0.2$, $b=1$ and $\alpha=0.8$.
The average seismicity rate is close to the expected value
$\mu^*= \mu /(n-1)$ predicted by (\ref{R1}) (dotted line)
and is always significantly larger than the constant external rate 
$\mu$ (dashed line). 78\% of earthquakes are aftershocks, among which
79\% are indirect aftershocks, in good agreement with
the predictions (\ref{R1}) and (\ref{r}).}
\end{figure}

\end{document}